\documentclass[11pt]{article}
\usepackage{amsfonts}
 \usepackage{amssymb}
  \def\bsh{\backslash}

 \newfont{\bbbold}{msbm10}

 \def\bbC{\mbox{\bbbold C}}

 \def\cF{{\cal F}}
 \def\cG{{\cal G}}

 \def\cO{{\cal O}}

 \def\cR{{\cal R}}

 \def\cV{{\cal V}}

 \newfont{\goth}{eufm10 scaled \magstep1}

 \def\gg{\mbox{\goth g}}

 \def\gl{\mbox{\goth l}}

 \def\gp{\mbox{\goth p}}

 \def\gs{\mbox{\goth s}}

 \def\ggl{\gg\gl}\def\gsl{\gs\gl}\def\psl{\gp\gs\gl}\def\pgl{\gp\gg\gl}
 
 \def\a{\alpha}

 \def\d{\delta}\def\D{\Delta}
 
 \def\f{\phi}

 \def\l{\lambda}

 \def\p{\pi}

 \def\th{\theta}

 \def\adt{\alpha'}

 \def\be{\begin{equation}}\def\ee{\end{equation}}
 \def\bea{\begin{eqnarray}}\def\eea{\end{eqnarray}}
 \def\ba{\begin{array}}\def\ea{\end{array}}

 \def\sd{\rm sdet}\def\str{\rm str}
 \def\xz{\times}

 \def\3dt{\dot{3}}
 
\def\half{{1\over2}}
 \let\la=\label

 \let\bm=\bibitem{}

 \def\bd{\begin{document}}
 \def\ed{\end{document}}
 \def\bea{\begin{eqnarray}}\def\barr{\begin{array}}\def\earr{\end{array}}
 \def\eea{\end{eqnarray}}
 \def\ft#1#2{{\textstyle{{\scriptstyle #1}\over {\scriptstyle #2}}}}
 \def\fft#1#2{{#1 \over #2}}
 \newcommand{\eq}[1]{(\ref{#1})}
 \def\eqs#1#2{(\ref{#1}-\ref{#2})}
 \def\det{{\rm det\,}}
 \def\tr{{\rm tr}}\def\Tr{{\rm Tr}}
  \def\str{{\rm str}} \def\diag{{\rm diag}}
 \def\sdet{{\rm sdet}}
\newcommand{\hoch}[1]{$^{#1}$}
 
\begin{document}

 \thispagestyle{empty}

 \hfill{KCL-MTH-02-26}

  \hfill{\today}

\vspace{20pt}

\begin{center}

{\Large\bf Four-point functions in $N=4$ SYM}

\vspace{30pt}

{\large P. J. Heslop\hoch1 and  P. S. Howe\hoch2}\\[20pt]

\end{center}

{\small \it \hoch1 Department of Physics,  Martin Luther Universit\"at, Halle, Germany}\\
{\small \it \hoch2 Department of Mathematics, King's College,
London, U.K. }

\vspace{100pt}

 {\bf Abstract}

A new derivation is given of four-point functions of charge $Q$
chiral primary multiplets in $N=4$ supersymmetric Yang-Mills
theory. A compact formula, valid for arbitrary $Q$, is given which
is manifestly superconformal and analytic in the internal bosonic
coordinates of analytic superspace. This formula allows one to
determine the spacetime four-point function of any four component
fields in the multiplets in terms of the four-point function of
the leading chiral primary fields. The leading term is expressed
in terms of $1/2 Q(Q-1)$ functions of two conformal invariants and
a number of single variable functions. Crossing symmetry reduces
the number of independent functions, while the OPE implies that
the single-variable functions arise from protected operators and
should therefore take their free form. This is the partial
non-renormalisation property of such four-point functions which
can be viewed as a consequence of the OPE and the
non-renormalisation of three-point functions of protected
operators.

 {\vfill\leftline{}\vfill \vskip  10pt

 \baselineskip=15pt \pagebreak \setcounter{page}{1}


\section{Introduction}

Four-dimensional superconformal theory has attracted renewed
interest in recent years mainly due to the Maldacena conjecture
which, in particular, relates IIB string theory on $AdS_5 \xz S^5$
to $N=4$ Yang-Mills theory \cite{maldacena}. In order to make
explicit checks of this conjecture it is necessary to try to make
computations on the field theory side of the correlation
functions. The non-renormalisation theorems for two- and
three-point functions of protected operators have been studied by
many authors (see \cite{agmo} for reviews and references}) and
have now been established non-perturbatively for all protected
operators \cite{Heslop:2001gp}.\footnote{It should be noted,
however, that there are operators with vanishing anomalous
dimensions which may not have non-renormalised three-point
functions.} It is therefore of interest to study four-point
functions, particularly those of the chiral primary operators
which correspond to the Kaluza-Klein states of IIB supergravity.
There has been a good deal of work on this topic too, as we review
briefly below. In this paper we discuss a new approach to these
correlation functions which is valid for arbitrary charges.

The main tool that we shall use is harmonic superspace, which was introduced
in a seminal paper in 1984 \cite{Galperin:1984av}; superspaces with
additional bosonic
coordinates were also discussed in a slightly different approach,
commonly referred to as projective superspace, in
\cite{Karlhede:vr}.
The maximally supersymmetric Yang-Mills theory in four spacetime
dimensions can be formulated in a rather neat way in $N=4$
analytic superspace \cite{Howe:1995md, Hartwell:1995rp}, an
example of a harmonic superspace. This
space has 8 even and 8 odd coordinates, compared to ordinary $N=4$
superspace which has 4 even and 16 odd coordinates. The additional
even coordinates describe a complex compact manifold which is a
coset space of the internal symmetry group $SU(4)$, while the
reduction in the number of odd coordinates can be viewed as a type
of generalised chirality. In this setting, the free $N=4$ SYM
field strength superfield is a single-component holomorphic field
$W$. The interacting field strength is covariantly analytic and so
is not itself a field on this space, but gauge-invariant
combinations of it are;  in particular, the so-called chiral
primary superfields $A_Q:=\tr (W^Q)$  are holomorphic fields on
analytic superspace (the gauge group is assumed to be $SU(N_c)$).
This fact was observed some time ago \cite{Hartwell:1995rp,hw2},
and a programme of investigating correlation functions of such
operators was initiated \cite{hw2,Howe:1998zi}.
In particular, a
lot of effort was expended on the analysis of the four-point
function of four CPOs, mainly in $N=2$ analytic superspace where
gauge-invariant products of the hypermultiplet are analytic
superfields.  Such four-point functions can be written in terms of
a prefactor multiplied by a function of the superconformal
invariants, and it was hoped that analyticity in the internal
coordinates would lead to restrictions on this function since the
natural superinvariants one can write down are all rational
functions of the coordinates. This is indeed the case, and the
full analysis was carried out in \cite{Eden:2000qp}.
 It is important to note
that
all of the unitary irreducible representations in $N=4$ can be realised as
superfields on analytic superspace, although, in general, these fields carry
superindices \cite{Heslop:2000mr,Heslop:2001zm}. This shows that the analytic superspace formalism is complete, and this is essential
for the study of OPEs and correlation functions as we shall see in this paper .

The basic result of \cite{Eden:2000qp} is that the $N=2$
four-point function of charge 2 operators can be solved in terms
of a single arbitrary function of the two independent spacetime
cross-ratios together with two other functions which satisfy two
coupled partial differential equations. These can be solved in terms
of two functions of single variables which can be constructed
from the cross-ratios \cite{Eden:2000bk}.
Explicit
calculations in perturbation theory \cite{Eden:1998hh,ess,Bianchi:2000hn} and in
the instanton sector \cite{Bianchi:1998nk} have been used to  find
particular examples of the two-variable function, and these
calculations also showed that the additional single-variable
functions take their free-theory forms.

A study of the four-point function of stress-tensor multiplets in $N=4$ was made
in \cite{Eden:2000bk}.  This is related to the charge 2 $N=2$ correlator because
$N=4$ SYM splits into a vector multiplet and a hypermultiplet in the $N=2$
formulation of the theory.  With the use of crossing symmetry and the reduction
formula it was shown that this four-point correlator is determined by a single
function of two-variables together with a part which is free\footnote{The
reduction formula relates the derivative of an $n$-point function with respect
to the coupling to an $(n+1)$-point function with one insertion of the
integrated on-shell action.  It was first used in the current context in
\cite{Intriligator:1998ig}}, and moreover agrees with the supergravity result \cite{dfmmr,dmmr,af}.
This is referred to as partial non-renormalisation
of the four-point function.  This result has recently been confirmed in a
component field OPE calculation where it was shown that the single-variable
functions arise from the interchange of protected operators \cite{Dolan:2000ut}.
The four-point function has also been studied in perturbation theory
\cite{Bianchi:1999ge,Bianchi:2000hn}, in
terms of instanton contributions \cite{Bianchi:1999ge,Bianchi:2000vh} and using the OPE
\cite{afp1,afp2,aeps,Dolan:2001tt}.

One implication of this result, first noted in \cite{afp1}, is that
some series A operators are
protected from developing anomalous dimensions even though it had been
widely
thought that they would be unprotected. This was confirmed in a
field theory calculation of three-point functions of two CPOs
and a third operator in \cite{Arutyunov:2001qw}; it can also be seen from
instanton contributions \cite{Bianchi:1999ge}.
Subsequently it was shown
that this phenomenon has a simple explanation and that there are
many more operators of this type:
the series A
operators which saturate unitarity bounds and which can be written
in $N=4$ analytic superspace in terms of derivatives and CPOs will
be guaranteed to satisfy the same shortening conditions as in the
free theory and will furthermore not be descendants
\cite{Heslop:2001dr}. Hence such operators will be protected
provided that the superconformal Ward identities are satisfied.

In this paper we revisit four-point functions of CPOs in the $N=4$
analytic superspace formalism. The key idea is to express the
function of invariants in terms of the Schur functions of the
group $GL(2|2)$. The group $GL(2|2)$ arises naturally in the
analytic superspace formalism as part of the isotropy subgroup of
the superconformal group which defines analytic superspace as a
coset (the analogous group in Minkowski space is the spin group
$SL(2)$). This group acts by the adjoint action on a
$(2|2)\xz(2|2)$ matrix $Z$ which is constructed from the four
coordinate matrices. The Schur functions are $\str\cR(Z)$ where
$\cR$ denotes a finite-dimensional representation of $GL(2|2)$.
Although we do not prove it here, this approach is related to the
OPE because one can identify the representations $\cR$ that occur
with the intermediate operators in the OPE (both primary and
descendants). The possible representations are restricted by
analyticity, and divide into two classes, the short, or atypical
representations and the long, or typical, representations. The
Schur functions are straightforward to compute, and combining this
information with crossing symmetry one arrives at compact
expressions for the four-point functions of four identical charge
$Q$ CPOs in terms of a number of functions of two conformal
invariants and a number of functions of only one variable. We
emphasise the fact that the formulae we give are valid for the
entire multiplets and not just for the leading scalar field
correlation functions. In principle one can compute the four-point
function of any four component fields in these multiplets in a
systematic fashion from the superfield formulae given in the
paper.

The functions of one variable are associated with the short
representations and the latter correspond to protected operators
in the OPE. Since the contribution of an intermediate operator is
related to the three-point function of two CPOs and the operator
in question, and since such three-point functions are
non-renormalised when the operator is protected, it follows that
the contribution of these operators should take the same form as
in the free theory. This is the partial non-renormalisation
theorem; from this point of view it can be seen as a corollary of
the existence of protected operators and the non-renormalisation
of three-point functions of protected operators.

In section 2  we review  the analytic superspace formalism for
$N=4$ SYM and then give a brief discussion of four-point functions
in $N=4$. In particular, we state the main result for the
contribution of long operators to such correlators. In section 3
we discuss Schur polynomials for various (super)groups of
interest. In section 4 we use the Schur functions to solve the
Ward identities for four-point functions of scalar fields or CPO
supermultiplets in four-dimensional theories with $N=0,2,4$
supersymmetries. These results are for four operators of the same
type but not necessarily identical. In section 5 we specialise to
the case of identical operators by imposing crossing symmetry. We
give formulae for the number of independent functions of two
variables which arise for different values of $Q$ in $N=2$ and
$N=4$.  Finally, in section 6 we discuss free four-point functions
for four charge 2 multiplets in both $N=2$ and $N=4$ and show how
these can be written in terms of functions of a single variable of
the type that arise in the preceding analysis.

\section{Four-point functions}

The field strength supermultiplet of $N=4$ super Yang-Mills theory
consists of six scalar fields transforming under the real
six-dimensional representation of $SU(4)$, four chiral spinor
fields and their conjugates and field strength tensor of the spin
one gauge field. In the free theory these fields can be packaged
into a single-component field $W(X)$ defined on analytic
superspace. This superspace has half the number of odd coordinates
as ordinary Minkowski superspace but also has an additional
compact internal space which can be viewed as the Grassmannian of
two-planes in $\bbC^4$ or as the coset space $S(U(2)\xz U(2))\bsh
SU(4)$. Locally this space is the same as complexified spacetime.
The local coordinates are

\be
X^{AA'} =\left( \ba{cc} x^{\a\adt} & \l^{\a a'} \\
\p^{a\adt} & y^{a a'} \ea \right) \ee

where the $x$s are the spacetime coordinates, the $y$s are local
coordinates on the internal space and the $\l$s and $\p$s are the
odd coordinates.  An infinitesimal superconformal transformation
acts on this space by

\be \d X^{AA'} =(\cV X)^{AA'} =B^{AA'} + A^A{}_B X^{BA'} +X^{AB'}D
_{B'}{}^{A'}+X^{A B'}C_{B' B}X^{B A'} \ee

where $\cV$ is the vector field generating the transformation and
where $A,B,C$ and $D$ are all $(2|2)\xz (2|2)$ supermatrices.  The
field $W$ transforms by

\be \d W=\cV W + \D W \ee

where

\be \D:=\str (A + XC)\ee

As it stands, this defines a $\pgl(4|4)$ transformation.
The Lie superalgebra $\psl(4|4)$ is the
complexification of the $N=4$ superconformal algebra and can be
obtained by imposing the constraint $\str (A + XC)=\str (D + CX)$,
while
$\pgl(4|4)$ extends $\psl(4|4)$ by an abelian factor
corresponding to the group $U(1)_Y$. It as first suggested
in \cite{Intriligator:1998ig} that $U(1)_Y$ could be a symmetry of the supercurrent
four-point function and this was confirmed in \cite{Eden:1999gh}.
In the following analysis we shall use the bigger group.

The chiral primary operators are defined to be single traces of
powers of $W$. In the interacting theory (we assume the gauge
group is $SU(N_c)$), $W$ is covariantly analytic and is not
actually a field on analytic superspace. However, gauge-invariant
products of the field strength are analytic fields, and so we can
define the CPOs as

\be A_Q:=\tr (W^Q) \ee

These operators transform in a similar way to $W$ but with a term
$Q\D A_Q$.

The four-point functions we are interested in are

\be <A_Q(X_1)A_Q(X_2)A_Q(X_3)A_Q(X_4)>:=<QQQQ> \ee

The superconformal Ward identities are

\be \sum_{i=1}^4 \left( \cV_i + Q \D_i\right) <QQQQ>=0 \ee

We can reduce the problem of solving these identities to that of
finding the superconformal invariants by introducing the
propagator $g_{12}$ which is the two-point function of $W$s in the
free theory:

\be <W(1)W(2)>\sim g_{12}=\sdet (X_{12}^{-1})={\hat y_{12}^2\over
x_{12}^2} \ee

where $X_{ij}:=X_i-X_j$ and where $\hat y=y-\p x^{-1} \l$. Here we
use the convention that inverse coordinates have lower indices,
e.g. $(x^{-1})_{\adt \a}$. With the aid of the propagator we can
write

\be <QQQQ>=(g_{12}g_{34})^Q \xz F_Q \ee

where $F_Q$ is a function of the superinvariants. A crucial
property of the four-point function is that it is analytic in the
internal coordinates (the $y$s). This must be so since each $A_Q$
is a polynomial in $y$ of degree $2Q$.

In previous analyses that have been made of the four-point
function in analytic superspace (mainly in $N=2$) the function $F$
has been written in terms of an explicit basis of invariants and
then the analyticity properties of functions of these has been
studied. In the current paper we adopt a different approach to
this problem by writing the invariants in terms of the Schur
polynomials of a certain variable $Z$.

Consider the problem of finding an invariant function $F$ of $n$
points $X_i$ in analytic superspace. It is clear that translation
invariance requires $F$ to be a function of difference variables
$X_{ij}$. Now define, for any three points $i\neq j\neq k$,

\be X_{ijk}:=X_{ij} X^{-1}_{jk} X_{ki} \ee

where no summations are implied. These transform by

\be \d X_{ijk}=A_i X_{ijk} + X_{ijk} D_i \ee

where $A_i:=A+X_iC$ and $D_i:=D+CX_i$. Provided that $F$ is
invariant under $A$ and $D$ transformations, invariance under
$C$-transformations will follow if $F$ is a function of $(n-2)$
variables of the type of $X_{ijk}$. We now define four-point
variables by

\be X_{ijkl}:=X_{ij} X^{-1}_{jk} X_{kl} X^{-1}_{li} \ee

where all points must be different. These variables transform by

\be \d X_{ijkl}=A_i X_{ijkl}-X_{ijkl} A_i \ee

As we shall show in more detail elsewhere, the $n$-point
invariants can be constructed from $(n-3)$ variables of this type
all of which transform under the adjoint  representation of
$\ggl(2|2)$ at point~1.

For four points there is only one independent such variable which
we shall call $Z$,

\be Z:=X_{2134}=X_{21} X^{-1}_{13} X_{34} X_{42}^{-1} \ee

The problem of four-point invariants is thus equivalent to finding
all functions of the matrix $Z$ which are invariant under the
adjoint action of  $\ggl(2|2)$.  The finite version of this is
invariance under the adjoint action of  the group $GL(2|2)$,

\be Z\mapsto G^{-1} Z G\ee

where $G\in GL(2|2)$. (Note that the full group does not act on
$Z$, only $PGL(2|2)$.) It can be shown that full invariants of
this type actually correspond to superconformal invariants which
are also invariant under $U(1)_Y$; for the case in hand, all
four-point invariants have this property. For $n\geq 5$ points one
can show that there are invariants which are invariant only under
$SL(2|2)$ rather than $GL(2|2)$ which give rise to superconformal
invariants which are not invariant under $U(1)_Y$.

The Schur polynomials are functions of the form $\str
(\cR(Z)):=S_{\cR}(Z)$ where $\cR$ denotes a finite dimensional
representation of $GL(2|2)$. Any such representation can be
described by a $GL(2|2)$ Young tableau labeled by four integers
$<m_1,m_2,m_3,m_4>$, but it is possible to restrict the types of
tableau that can occur if we also allow powers of the
superdeterminant of $Z$.  We thus arrive at the formula

\be <QQQQ>=(g_{12} g_{34})^Q \sum_{p,\cR} C_{p,\cR} \ (\sdet Z)^p
S_{\cR} (Z)\la{eq:16} \ee

This formula is related to the OPE approach to four-point
functions. The OPE for two $A_Q$s was given
in~\cite{Heslop:2001gp}; it reads

\be A_Q(1) A_Q(2) \sim \sum_{\cR} (g_{12})^{Q-{q\over2}}
\cR(X_{12})\cO^{q}_{\cR\cR}(2) + \ldots \ee

Here, the dots denote the contributions of the descendants of the
primary fields $\cO^q_{\cR\cR}$ and $\cR(X_{12})$ means a product
of $X_{12}$ with both the primed and the unprimed indices
projected into the representation $\cR$. Each primary field in
this expansion has charge $q=L-(J_1+J_2)$ where $L$ is the
dilation weight and $J_1, J_2$  are the two spin quantum numbers
of the superconformal representation under which the operator
transforms. In general, an operator on analytic superspace will be
a tensor (or quasi-tensor) field carrying $I$ $A$-type
superindices and an equal number of primed superindices and will
transform under finite-dimensional irreducible representations of
the two $GL(2|2)$ groups which act on the primed and unprimed
superindices $A,A'$.  The number of indices of each type must be
equal in order to have vanishing R-weight, and all of the indices
must be covariant (subscript) in order for the representation to
be unitary.  Since the two operators on the left-hand-side of the
OPE are scalars,  operators contributing to the right-hand-side
will have $\cR=\cR'$. Analyticity places restrictions on the
representations that can appear in the OPE, but these can be seen
more easily directly from analyticity of the four-point function.
A key restriction imposed by unitarity is that one only sees
representations $\cR$ corresponding to covariant tensorial
representations. The conjugate representations to these will
therefore not appear in the expansion of the four-point function.

The contribution of an operator $\cO^{q}_{\cR\cR}$ to the
four-point function $<QQQQ>$ has the form

\be <QQQQ> \sim {(A_{QQ\cO} )^2\over A_{\cO\cO}} (g_{12} g_{34})^Q
(\sdet Z)^{q/2}\sum_{\cR'}  C_{\cR'} S_{\cR'} (Z)\la{eq:18} \ee

where $\cR'$ denotes a  representation with a Young tableau which
contains the Young tableau of $\cR$, $A_{QQ\cO}$ is the
coefficient of the 3-point function $<A_QA_Q\cO>$, $A_{\cO\cO}$ is
the coefficient of the 2-point function $<\cO\cO>$, and $C_{\cR'}$
are purely numerical constants (in particular $C_{\cR}=1$.)

It is simple to obtain the restrictions that analyticity imposes
on $F_Q$. From the form of $Z$ it is apparent that there can only
be poles in the $(13)$ and $(24)$ channels from each $S_{\cR}(Z)$.
Looking at the $(12)$ channel we have $g_{12}^Q\sim (y_{12}^2)^Q$
while $\sdet (Z)^p\sim (y_{12}^2)^{-p}$. Hence analyticity in this
channel requires $p\leq Q$. The same result is obtained from the
$(34)$ channel.  For the $(13)$ channel $\sdet Z^p\sim
(y_{13}^2)^p$ while the leading singularity from $S_{\cR}(Z)$ will
arise from the term with most factors of $y$. As we shall see
below,  the tableaux we need to consider are those which have two
rows of arbitrary length (provided that the second is not longer
than the first) together with a single first column of length $r$
or the trivial representation which has $r=0$. The worst
singularity one can have for all of these possibilities is
$(y_{13}^2)^{-r}$ from which we learn that $p\geq r$.  The same
result is obtained from the $(24)$ channel. So the restrictions
due to analyticity are simply

\be r\leq p\leq Q. \ee

In order to write the correlator in a more explicit form it is
useful to use the $G$ transformation of $Z$ to bring it diagonal
form. We can thus write it in terms of its eigenvalues as

\be Z\sim \diag (X_1,X_2| Y_1,Y_2) \ee

In section~(\ref{sec:schur}) we shall give explicit formulae for
the Schur polynomials in terms of these variables, but for the
moment we shall simply state that they can be used to show that
the contribution of the long operators to the correlator, can be
written in the form

\be <QQQQ>_2= (g_{13}g_{24})^Q S F^Q(X_1,X_2,Y_1Y_2)
\la{mainresult} \ee

where $S$ is a universal function of the eigenvalues which will be
given explicitly later on. $F^Q$ is a polynomial in the variables
$(Y_1+Y_2)$ and $Y_1Y_2$ of degree $Q-2$ with coefficients which
depend on the variables $X_1,X_2$. Now these variables, evaluated
at zero in the odd coordinates, will simply be the eigenvalues of
the $GL(2)$ matrix $z=x_{21}x_{13}^{-1}x_{34}x_{42}^{-1}$ which
occurs in the bosonic problem, so that these coefficient functions
can equivalently be regarded as functions of the two independent
four-point conformal invariants in spacetime. At this stage,
therefore, \eq{mainresult} tells us that the four-point function
of CPOs is determined completely (to all orders in odd coordinates)
by $1/2\ Q(Q-1)$ functions of two variables, together with a part
which describes the contribution of the protected operators.
However, we can also impose crossing symmetry as we have taken the
CPOs to be identical. Under the interchange of point 1 and 3
$Z\mapsto (1-Z)$,  while the prefactor $(g_{13}g_{24})^Q S$ is
invariant. Under the interchange of points $1$ and 4 $Z\mapsto
Z^{-1}$ while the prefactor changes to itself times
$(Y_1Y_2)^{Q-2} (X_1 X_2)^{-Q-2}$. Demanding symmetry under these
two operations is sufficient to ensure full crossing symmetry and
reduces the number of independent functions considerably. For low
values of $Q$, we find that the number of independent functions is
1 for $Q=2$ and for $Q=3$, 2 for $Q=4$, 3 for $Q=5$ and so on. We
shall discuss crossing symmetry in more detail in section 5 where
we give a formula for the number of independent functions for
arbitrary $Q$.

To summarise, the four-point function of four identical CPOs of
charge Q is given by \eq{mainresult} together with the
contribution from protected operators. The function $F^Q$ is
determined by a number of functions of two variables which depends
on $Q$. This result generalises the partial non-renormalisation
theorem for the supercurrent multiplet (which corresponds to
$Q=2$). The partial non-renormalisation concerns the contribution
of the protected operators via the OPE. As we shall see below,
these operators give rise to functions of a single variable (i.e.
one of the eigenvalues $X_1$ or $X_2$). Since these functions are
related to the three-point functions of two $A_Q$s and the
protected operator in question, it follows that they are
non-renormalised. We thus expect these single-variable functions
to take the same form in the interacting theory as they would in
the free theory.

\section{Schur polynomials}\la{sec:schur}

As we have seen above a four-point invariant function $F$ can be
written entirely as a function of the $(2|n) \xz (2|n)$ matrix
$Z=X_{21} X_{13}^{-1}X_{34} X_{42}^{-1}$ which is invariant under
the adjoint action, $Z\rightarrow {G}^{-1} Z {G}$ for any
$GL(2|2)$ matrix $G$. In fact we can consider the more general
case of $GL(2|n)$, since then $n=0$ corresponds to Minkowski
space, $n=1$ to $N=2$ analytic superspace and $n=2$ to $N=4$
analytic superspace.

Such an invariant function is known as a class function. It is
useful to write such functions in terms of the Schur polynomials
or characters:

\be S_{\cR}(Z):=\str \cR(Z). \ee

where $\cR$ is any representation of $G$. Indeed,  for Lie groups,
the Peter-Weyl theorem states
 that the characters of irreducible
 representations span a dense subspace of the space of continuous class
 functions (see~\cite{FH}).

In practice the Schur polynomials can be obtained as follows. The
representation $\cR$ is specified by a Young tableau with, say $m$
boxes. The expression $\cR(Z)$ is obtained by taking the tensor
product of $m$ $Zs$ and symmetrising the upstairs indices
according to the Young tableau. The supertrace over this
representation is then obtained by contracting the upstairs and
downstairs indices, and dividing by the hook length formula for
the Young tableau in question\footnote{The hook length formula is
the product of the hook lengths of all the boxes in a Young
tableau, the hook length of a box being the number of squares
below and to the right of the box, including the box itself
once.}. So for example for the fundamental representation, $\cR=
\square$ one has simply the usual supertrace of $Z$,

\be \cR=\square \Rightarrow S_{\cR}(Z)=\str(Z),\la{fund} \ee

whereas for the symmetric representation one has

\be \cR= \square \! \square \Rightarrow S_{\cR}(Z)=\half (Z^A{}_A
Z^B{}_B (-1)^{A+B}+ Z^B{}_A Z^A{}_B (-1)^B)=\half(\str(Z)^2 +
\str(Z^2)).\la{sym} \ee

since the Hook-length formula gives $2$ in this case. We shall
expand the invariant functions we are interested in in terms of
these Schur polynomials. The assumption that this is possible can
be shown to be correct for four-point functions which have a
double OPE expansion interpretation.

\subsection{Schur polynomials in terms of invariant variables}

A useful way to write the Schur functions explicitly is in terms
of the
 eigenvalues of $Z$. To do this one chooses a matrix $G$ which
diagonalises $Z$
 so
that

\be Z=\mbox{diag}(X_1,X_2|Y_1,\dots Y_n). \ee

Since the 4-point invariant function $F$ is invariant under the
adjoint action, it can be written entirely as a function of the $2
+n$ eigenvalues of $Z$,

\be F=F(X_1,X_2,Y_1\ldots Y_n). \ee

It is straightforward, in principle, to find these variables
although, in practice, it is in general a hard problem to find the
matrix $G$ which diagonalises $Z$. In addition, because there are
matrices $G$ which can interchange $X_1, X_2$ and also
 (separately)
permute the $Ys$, the function F must be a symmetric function of
$X_1,X_2$ and  of the $Y_i $s.

To make things more explicit we now express the Schur polynomials
in terms of the eigenvalues of $Z$. We consider several different
cases in turn.

\subsubsection{$ GL(1)$}

First, as a warm up, let us consider the trivial case $GL(1)$ in
which case $Z$ is simply a complex number. Representations of
$GL(1)$ are given by Young tableaux with only one row with $n$
boxes say. In this case one simply finds that

\be
 S_{\cR}(Z)=Z^n
 \ee

One must also consider conjugate representations, obtained by
replacing $Z$ by $Z^{-1}$. These can be accommodated by allowing
$n$ to take
 take negative
values in the above formula.

\subsubsection{$ GL(2)$}

This corresponds  to ordinary four-dimensional conformal  symmetry
(i.e. $N=0$).
 In this case we shall write the diagonal form of $Z$ as $Z=\diag(x_1,x_2)$
and so
 the Schur
 polynomials can be written in terms of these two variables.
 Representations of $GL(2)$ are given by Young tableaux
 with two rows of lengths $a+b$ and $b$ say, and the corresponding Schur
 polynomials are given as:

 \be \cR_{ab}=\begin{picture}(100,0) \put(0,0){$\overbrace{\square \!
\square  \cdots \square \! \square}^b \! \!
 \overbrace{\square \! \square \cdots  \square \! \square}^a $}
\put(0,-7){$ \square \! \square  \cdots \square \! \square  $}
\end{picture}
\qquad \Longrightarrow  \qquad S_{ab}(Z)=(x_1x_2)^b
\left({x_1^{a+1}-x_2^{a+1} \over x_1-x_2} \right) \la{schur}\ee
\\

Again, one should also consider conjugate representations.
However, $GL(2)$ representations are equivalent to their conjugate
representations up to multiplication of a determinant. In fact,
all representations can be included by allowing $b$ to take
negative values in the above formula. However, an OPE analysis
together with CFT unitary bounds shows that the conjugate
representations are not needed in the four-point function. Note
that since the conformal group is non-compact $b$ may also take
non-integer values
 corresponding to the presence of operators with anomalous dimensions.

We briefly note here that the variables $x_1,x_2$ are precisely
the same as the variables $x,z$ used in~\cite{Dolan:2000ut}; they
are related to the usual cross-ratios $u,v$ by~\eq{uv}.

We shall want to consider arbitrary linear combinations of the
Schur polynomials; these can be rewritten in terms of a function
of two variables as follows:

\be \sum_{a,b} C_{ab} S_{ab}(Z)={G(x_1,x_2)}\la{f2} \ee

where

\be
 G(x_1,x_2):= \sum_{a,b} C_{ab}\  (x_1x_2)^{b} \ {x_1^{a+1}-x_2^{a+1}\over
    x_1-x_2}.
\ee

and where $C_{ab}$ are arbitrary constants.

\subsubsection{$ GL(2|1)$}

 The case $GL(2|1)$ corresponds to $N=2$ superconformal symmetry.
 The easiest way to find the Schur polynomials for supergroups is
 to consider first the representations as representations of the maximal
 bosonic
 subgroup,
 in this case $GL(2|1)\supset GL(2)\xz GL(1)$. Splitting $GL(M|N)$
 representations into $GL(M) \xz GL(N)$ is done in a similar way
 as for $GL(M+N)$ except that one considers the conjugate
 representation of $GL(N)$ (for more detail on this
see~\cite{Bars:1983se}.)

 All representations of $SL(2|1)$ can be given by Young tableaux
 with two rows, and hence representations of $GL(2|1)$ can be
 given by Young tableaux with two rows up to multiplication by
 the superdeterminant.
 The representations come in two types, $R_{ab}$ and $R_{a}$, which
 have Young tableaux and $\gsl(2|1)$ Dynkin labels as follows:

 \bea
 \cR_{ab}=\begin{picture}(100,0)
\put(0,0){$\overbrace{\square \! \square  \cdots \square \!
\square}^{b-1} \! \!
 \overbrace{\square \! \square \cdots  \square \! \square}^a $}
\put(0,-7){$ \square \! \square  \cdots \square \! \square  $}
\end{picture}&\qquad& [a, a+b] \quad b>1,\ a\geq0 \la{N=2reps}\\
\cR_{a}= \overbrace{\square \! \square \cdots \square \!
\square}^{a+1}&\qquad& \left\{ \ba{lcl} [a, a+1] &\quad& a\geq0\\
      \left[ 0, 0 \right]    &\quad& a=-1\ea
                                        \right.
 \eea

The $\cR_{ab}$s are long (or typical) representations whereas the
$\cR_a$s are short (or atypical). Note that $a$ must be an integer
while $b$ can be non-integral. In the context of superconformal
field theory $b$ must be real and greater than 1 and the presence
of non-integral representations corresponds directly to anomalous
dimensions. We remark that the interpretation of a Young diagram
is the same as in the bosonic case with regard to symmetrisation,
except for the fact that symmetry or anti-symmetry is understood
to be generalised. Thus, for example, $\cR_1$ corresponds to a
tensor which is generalised symmetric on two $\bbC^{2|1}$ vector
indices.

 Under $GL(2|1)\supset GL(2)\xz GL(1)$ we find

 \bea
\cR_{a}= \overbrace{\square \! \square \cdots \square \!
\square}^{a+1} & \rightarrow & (\overbrace{\square \! \square
\cdots \square \! \square}^{a+1},1 ) + ( \overbrace{\square \!
\square \cdots \square \!
\square}^{a},\square) \\
\cR_{02}=\begin{picture}(10,10)(-2,-3) \put(0,0){$\square $}
\put(0,-7){$ \square  $}
\end{picture}
& \rightarrow & \left( \begin{picture}(10,10)(-2,-3)
\put(0,0){$\square $}\put(0,-7){$ \square $}\end{picture},1
\right)
 +
 \left( \square ,\square
\right)
 +
  \left( 1,\square \! \square \right).
  \eea

 The Schur polynomials are just the supertraces over these
 representations of $Z \sim \diag (X_1,X_2|Y)$, so we can write
 the Schur polynomials of the supergroup in terms of the corresponding
Schur
 polynomials of the maximal bosonic subgroup $GL(2)\xz GL(1)$, in terms
of the
 variables
 $X_1,X_2$ for $GL(2)$ and $Y$ for $GL(1)$. So for example,
 corresponding to the above two representations we have the Schur
 polynomials

\bea
 S_a(Z)&=&{X_1^{a+2}-X_2^{a+2} \over X_1-X_2} -Y{(X_1^{a+1}-X_2^{a+1})
\over
 X_1-X_2} \qquad
 a\geq-1 \la{S_a}\\
 S_{02}(Z)&=&X_1X_2 -Y (X_1+X_2) + Y^2.
 \eea

 Notice that a minus  sign occurs whenever there are an odd number of
 $Y$'s. This is
 because of the minus sign in the definition of the supertrace.

 To find the Schur polynomials for more general representations
 with two rows, notice that under $GL(2|1)\supset GL(2) \xz GL(1)$

 \be
 \cR_{ab}=\begin{picture}(100,0)
\put(0,0){$\overbrace{\square \! \square  \cdots \square \!
\square}^{b-1} \! \!
 \overbrace{\square \! \square \cdots  \square \! \square}^a $}
\put(0,-7){$ \square \! \square  \cdots \square \! \square  $}
\end{picture}
 =
 \left\{ {\left( \begin{picture}(10,10)(-2,-3)
\put(0,0){$\square $}\put(0,-7){$ \square $}\end{picture},1
\right)
 +
 \left( \square ,\square
\right)
 +
  \left( 1,\square \! \square \right) }\right\}
  \xz
  \left( {\begin{picture}(100,0)
\put(0,0){$\overbrace{\square \! \square  \cdots \square \!
\square}^{b-2} \! \!
 \overbrace{\square \! \square \cdots  \square \! \square}^a $}
\put(0,-7){$ \square \! \square  \cdots \square \! \square  $}
\end{picture},1
 }\right)
\ee

 and since the Schur polynomials respect the multiplication of
 representations (i.e. $S_{\cR} S_{\cR'}=S_{\cR \cR'}$) we find that

 \be
 S_{ab}(Z)=(X_1X_2)^{b-2} \left({X_1^{a+1} -X_2^{a+1} \over
X_1-X_2}\right)
 S_{02}(Z).
 \ee

As in the purely bosonic case we will consider linear combinations
of Schur polynomials and rewrite them in terms of functions as
follows

\be \sum_a C_a S_a ={ X_1f(X_1)-X_2f(X_2) \over X_1-X_2}-Y \left(
{f(X_1)-f(X_2)
 \over X_1-X_2}
 \right)  :=\cF[f] \la{Ff}
\ee

where

\be f(X)= \sum_a C_a X^{a+1}\la{1v} \ee

and

\be \sum_{a,b} C_{ab} S_{ab} = {G(X_1,X_2)}(X_1X_2 -Y (X_1+X_2) +
Y^2) := \cG[G] \la{Gg} \ee

where

\be G(X_1,X_2):=\sum_{a,b}C_{ab} \ (X_1X_2)^{b-2}\
{X_1^{a+1}-X_2^{a+1}\over X_1-X_2}.\la{2v} \ee

We see that the short representations lead to a function of one
variable, $f(X)$ whereas the long representations lead to a
function of two variables $G(X_1,X_2)$. These are the origin of
the one and two variable functions
 obtained
in the analysis of 4-point functions as we shall see shortly.

The long representations are a priori only valid for $b>1$
although $b$ can take non-integer values corresponding to the
presence of anomalous dimensions and the related notion of
quasi-tensors~\cite{Heslop:2001gp}). However, we may consider the
formula for $S_{ab}$ for the special case $b=1$,
 $S_{a1}$.
According to~\eq{N=2reps} this appears to correspond to the
representation with
 Dynkin labels $[a,a+1]$ in other words
$\cR_{a1}$ has the same Dynkin labels as $\cR_a$. However, these
two are not the same as can be seen by comparing the Schur
polynomials $S_{a1}$ and $S_{a}$. The reason for this apparent
discrepancy is that $\cR_{a1}$ is a reducible representation where
as $\cR_a$ is irreducible. This can be seen from the corresponding
Schur polynomials which satisfy

\be S_{a1}(Z)= S_{a-1}(Z) -\sdet(Z)^{-1} S_{a} (Z) . \la{red} \ee

So $\cR_{a1}$ contains the representations $R_a$ and
$R_{a-1}$\footnote{This
  result is directly related to the reducibility of certain
superconformal
  operators at the threshold of the unitary bounds, such as the Konishi
  multiplet.}.

Note that the $N=2$ superconformally invariant variables $X_1,X_2$
are related to the conformally invariant variables $x_1,x_2$ by
$X_1=x_1 +O(\th), X_2=x_2+O(\th)$ where $\th$ are Grassmann odd
coordinates.

In principle, we should also include the conjugate representations
which in the supersymmetric case are not equivalent to
non-conjugate representations. However, an analysis of the OPE in
conjunction with the superconformal unitary bounds shows that the
conjugate representations are not needed for 4-point functions.

\subsubsection{$ GL(2|2)$}

The $GL(2|2)$ case, corresponding to $N=4$ superconformal
symmetry, is similar to the $GL(2|1)$ case. We shall illustrate
the procedure for finding Schur polynomials with two examples, and
then just write them down for general representations.

 In this case there are again two types of representation which we denote
 $\cR_{abc},\ \cR_{ac}$ with Young tableaux and Dynkin
 labels

 \be
 \begin{picture}(300,70)(0,-50)
   \put(-50,0){$\cR_{abc}=$}\put(0,0){$\overbrace{ \square
\! \square  \cdots \square \! \square}^b \! \!
 \overbrace{\square \! \square \cdots  \square \! \square}^a $}
\put(0,-7){$ \square \! \square  \cdots \square \! \square  $}
\put(-14.7,-30.4){$\left\{ \ba{c} \square \\ \vdots \\ \square \ea
\right. $} \put(-20,-30){$\scriptstyle c$}
\put(120,0){$[a,a+b,c],\ b>1,\ a,c\geq0 $}\end{picture}
 \ee
 \be
 \begin{picture}(300,50)(0,-40) \put(-50,-7){$\cR_{ac}=$}
\put(0,-7){$ \overbrace{\square \! \square  \cdots \square \!
\square }^{a+2} $} \put(-14.7,-30.4){$\left\{ \ba{c} \square \\ \vdots \\
\square \ea \right. $} \put(-20,-30){$\scriptstyle c$}
\put(100,-27){$\ba{ll} [a, a+1, c]& a,c\geq0\\
\left[0,0,c+1\right]&
  a=-1,c\geq0 \ea$}
\end{picture}
\ee

The trivial representation must be treated seperately in this case
and we shall call it $\cR_0$. For these representations $a$ and
$c$ are integral while $b$ can again be non-integral. As in the
$GL(2|1)$ case, to find the Schur polynomials for these
representations, we begin by splitting the supergroup into its
maximal bosonic subgroup. For example, under $GL(2|2) \supset
GL(2) \xz GL(2)$

\bea \cR_{a0}= \overbrace{\square \! \square \cdots \square \!
\square}^{a+2} & \rightarrow & (\overbrace{\square \! \square
\cdots \square \! \square}^{a+2},1 ) + ( \overbrace{\square \!
\square \cdots \square \! \square}^{a+1},\square)+ (
\overbrace{\square \! \square \cdots \square \!
\square}^{a},\square \! \square ) \\
\cR_{020}=\begin{picture}(17,10)(-2,-3) \put(0,0){$\square \!
\square \! \! $} \put(0,-7){$ \square \! \square $}
\end{picture}
& \rightarrow & \left( \begin{picture}(17,10)(-2,-3)
\put(0,0){$\square \! \square \! \! $}\put(0,-7){$ \square \!
\square $}\end{picture},1 \right)
 +
 \left( \begin{picture}(17,10)(-2,-3) \put(0,0){$\square \!
\square \! \! $}\put(0,-7){$ \square $}\end{picture},\square
\right)
 +
  \left( \begin{picture}(10,10)(-2,-3) \put(0,0){$\square
 $}\put(0,-7){$ \square
$}\end{picture},\square \! \square \right)
 +
 \left( \square \! \square,\begin{picture}(10,10)(-2,-3)
\put(0,0){$\square
 $}\put(0,-7){$ \square
$}\end{picture} \right)
 +
 \left( \square,\begin{picture}(17,10)(-2,-3) \put(0,0){$\square \!
\square \! \! $}\put(0,-7){$ \square $}\end{picture} \right)
 +
 \left( 1,\begin{picture}(17,10)(-2,-3)
\put(0,0){$\square \! \square \! \! $}\put(0,-7){$ \square \!
\square $}\end{picture} \right) \eea

and the corresponding Schur polynomials, given in terms of $Z\sim
\diag(X_1,X_2,Y_1,Y_2)$ are

\be  S_{a0}(Z) = {X_1^{a+3}-X_2^{a+3}\over
X_1-X_2}-(Y_1+Y_2){X_1^{a+2}-X_2^{a+2}\over X_1-X_2}+Y_1 Y_2
 {X_1^{a+1}-X_2^{a+1}
\over X_1-X_2} \quad  a\geq -1 \la{c=0}\ee

\be \ba{rcl} S_{020}(Z)&=& (X_1X_2)^2-X_1X_2(X_1+X_2)(Y_1+Y_2)
+Y_1 Y_2 \left({X_1^3-X_2^3 \over
X_1-X_2}\right)\\
&+&  \left({Y_1^3-Y_2^3 \over Y_1-Y_2}\right) X_1X_2 - Y_1 Y_2(Y_1
+Y_2)(X_1+X_2) +(Y_1 Y_2)^2. \ea \la{S020}\ee

Note that this time both the space-time group and the internal
group is $GL(2)$.

Note also that the $N=4$ variables $X_1,X_2$ are not of course the
same as in
 the
$N=2$ case. It should be clear from the context whether we are
talking about $N=4$ variables or $N=2$ variables.

By decomposing the representations of $GL(2|2)$ into
representations of $GL(2)\xz GL(2)$ in this way one finds the
following Schur polynomials:

 \bea
 &&S_{abc}=(-1)^{c}(X_1X_2)^{b-2} \left({X_1^{a+1} -X_2^{a+1} \over
 X_1-X_2}\right)\left({Y_1^{c+1}-Y_2^{c+1} \over Y_1-Y_2}\right) \xz
 S_{020}
  \la{N=4}\\
&&\ba{rcl} S_{a,c}&=& (-1)^{c+1}(X_1X_2) \left( {X_1^{a+2}
-X_2^{a+2} \over X_1-X_2} \right)\left({Y_1^{c}-Y_2^{c} \over
Y_1-Y_2}\right)\\
 &+&
 (-1)^{c}(X_1+X_2) \left( {X_1^{a+2} -X_2^{a+2}
\over X_1-X_2} \right)\left({Y_1^{c+1}-Y_2^{c+1} \over
Y_1-Y_2}\right)\\
 &+&
(-1)^{c}(X_1X_2) \left( {X_1^{a+1} -X_2^{a+1} \over X_1-X_2}
\right)(Y_1 Y_2)\left({Y_1^{c-1}-Y_2^{c-1} \over
Y_1-Y_2}\right) \qquad \qquad (c\geq 0, a\geq-1)\\
 &+&
 (-1)^{c+1}\left( {X_1^{a+2} -X_2^{a+2}
\over X_1-X_2} \right)\left({Y_1^{c+1}-Y_2^{c+1} \over
Y_1-Y_2}\right)(Y_1+Y_2)\\
 &+&
 (-1)^{c+1}(X_1X_2) \left( {X_1^{a} -X_2^{a}
\over X_1-X_2} \right)(Y_1 Y_2)\left({Y_1^{c}-Y_2^{c} \over
Y_1-Y_2}\right)\\
 &+&
(-1)^{c}\left( {X_1^{a+1} -X_2^{a+1} \over X_1-X_2}
\right)(Y_1Y_2)\left({Y_1^{c+1}-Y_2^{c+1} \over Y_1-Y_2}\right).
\ea \la{S_ac}\eea

If we analytically continue $S_{ac}$ down to the case $a=-2,\
c=0$, which should correspond to the trivial representation, we
obtain instead $S_{-2,0}=1-Y_1Y_2/X_1X_2$, whereas for the trivial
representation we should obtain the answer $S_{0}=1$, so we treat
this case separately.

We will again consider linear combinations of these Schur
polynomials. These can be expressed in terms of functions of one
or two variables:

\be \sum_{a,b}C_{abc}S_{abc}({Z})= (-1)^c  G_c(X_1,X_2)
 \left({Y_1^{c+1}-Y_2^{c+1} \over Y_1-Y_2}\right) \xz
 S_{020}({Z}):=\cG_c[G_c] \la{eq:52}
\ee

where

\be
 G_c(X_1,X_2)= \sum_{a,b} C_{abc} (X_1X_2)^{b-2}\  {X_1^{a+1}- X_2^{a+1}\over
    X_1-X_2}
\ee

and

\be \ba{rcl} \sum_a C_{ac} S_{ac}&=& (-1)^{c+1}(X_1X_2) \left(
{X_1^2f_c(X_1)
 -X_2^{2}f_c(X_2)
  \over
X_1-X_2} \right)\left({Y_1^{c}-Y_2^{c} \over
Y_1-Y_2}\right)\\
 &+&
 (-1)^{c}(X_1+X_2) \left( {X_1^2f_c(X_1) -X_2^{2}f_c(X_2)
\over X_1-X_2} \right)\left({Y_1^{c+1}-Y_2^{c+1} \over
Y_1-Y_2}\right)\\
 &+&
(-1)^{c}(X_1X_2) \left( {X_1f_c(X_1) -X_2f_c(X_2) \over X_1-X_2}
\right)(Y_1 Y_2)\left({Y_1^{c-1}-Y_2^{c-1} \over
Y_1-Y_2}\right) \\
 &+&
 (-1)^{c+1}\left( {X_1^{2}f_c(X_1) -X_2^{2}f_c(X_2)
\over X_1-X_2} \right)\left({Y_1^{c+1}-Y_2^{c+1} \over
Y_1-Y_2}\right)(Y_1+Y_2)\\
 &+&
 (-1)^{c+1}(X_1X_2) \left( {f_c(X_1) -f_c(X_2)
\over X_1-X_2} \right)(Y_1 Y_2)\left({Y_1^{c}-Y_2^{c} \over
Y_1-Y_2}\right)\\
 &+&
(-1)^{c}\left( {X_1f_c(X_1) -X_2f_c(X_2) \over X_1-X_2}
\right)(Y_1Y_2)\left({Y_1^{c+1}-Y_2^{c+1} \over
Y_1-Y_2}\right):=\cF_c[f_c] \ea \ee

where

\be f_c(X_1)=\sum C_{ac}X_1^a. \ee

Once again we see that the short representations of $GL(2|2)$ lead
to
 functions of one variable, whereas the long representations lead to
functions
 of two variables.

 Formula~\eq{N=4} is a priori only valid for $b>1$. However, as in the
$N=2$
 case it is also possible to use the formula for $b=1$, in
 which case it corresponds to a reducible representation.
 In fact one finds that

\be S_{a1c}(Z)= S_{(a-1) (c+1)}(Z) +\sd(Z)^{-1}S_{ac} (Z).
\la{red2} \ee

giving the analogue of \eq{red} for $N=2$.

 For the same reason as in the $N=2$ case we do not need to include the
 conjugate representations of $GL(2|2)$.

\subsection{Relation to other variables}

We may now ask how these relate  these variables to those used
previously in various papers. In~\cite{Dolan:2000ut} it was found
useful to consider the variables $x_1,x_2$ (called
 $x,z$ in the paper)
where

\be x_1 x_2= u:= {x^2_{12} x^2_{34} \over x^2_{13} x^2_{24}}
\qquad (1-x_1)(1-x_2)=v:= {x^2_{14} x^2_{23} \over x^2_{13}
x^2_{24}}.\la{uv} \ee

Interestingly enough, the eigenvalues of $Z$ which we are using,
$x,z$, are precisely these variables in \cite{Dolan:2000ut}. To
see this note that $u=\mbox{det}(Z)=xz$ and $1-Z= X_{23}
X_{31}^{-1}X_{14} X_{42}^{-1}$ which means that
$v=\mbox{det}(1-Z)=(1-x)(1-z)$ giving exactly the
relations~\eq{uv}. This elucidates the importance of the variables
$x,z$. (We have called these $(x_1,x_2)$ above.)

For $n=1$ (i.e. N=2 superconformal field theory) the invariant
variables

\be V={T+U-1 \over 1+S-T} \qquad S'=SV \qquad T'=T(1+V) \ee where
\be S={\sdet {X_{14}}\, \sdet {X_{23}} \over \sdet { X_{12}
}\,\sdet { X_{34}} }\qquad T={\sdet { X_{13} }\,\sdet { X_{24}
}\over \sdet { X_{12} }\,\sdet { X_{34}} }\qquad U=\str
(X_{12}^{-1}X_{23}X_{34}^{-1}X_{41}) \ee

were found to be useful in~\cite{Eden:2000qp}. These can be
straightforwardly related to the eigenvalues of $Z$,  i.e.
$X_1,X_2,Y$. It is perhaps more illuminating to consider instead
the matrix $Z'=X_{23} X_{34}^{-1}X_{41} X_{12}^{-1}$ (this is
straightforwardly related to $Z$ by permuting the insertion
points) and its corresponding eigenvalues $X'_1,X'_1,Y'$. Then one
finds straightforwardly that

\be S=-{X'_1X'_2 \over Y'}\qquad T={(1-X'_1)(1-X'_2)\over (1-Y')}
\qquad
 U=X'_1+X'_2-Y'
\ee

and that \be S'=X'_1X'_2 \qquad T'=(1-X'_1)(1-X'_2) \qquad V=-Y'
\ee

Thus we see that $V$ is (up to a minus sign) the same as the
internal eigenvalue $Y$ whereas $S',T'$ have a similar
relationship to $X'_1,X'_2$ as $u,v$ do to $x,z$ (compare with
equation~\eq{uv}.)

\section{Construction of four-point functions}

In this section we consider the four-point functions of various
gauge invariant operators in (super)conformal field theories with
the aid of the representation theory of $GL(2|n)$. Consideration
of the OPE expansion~\eqs{eq:16}{eq:18} tells us that the
four-point function of charge $Q$ scalar operators $A_Q$ can be
expanded in the form

\be <QQQQ>=(g_{12} g_{34})^Q \sum_{p,\cR} C_{p,\cR}  \ \sdet
(Z)^{p} \ S_{\cR}(Z) \la{4ptexp}\ee

where $p\geq0$, $C_{p,\cR}$ are arbitrary constants and
$S_{\cR}(Z)$ are Schur polynomials with positive powers of the
eigenvalues only (i.e. we do not need conjugate representations).

Before considering analytic superspace itself, we consider two
simpler cases for comparison, namely $GL(2)$, which corresponds to
the component formalism for conformal field theories, and
$GL(0|n)$ which gives some insight into the internal $SU(2n)$
symmetry of superconformal field theory.

\subsection{Four-point functions in $N=0$}\la{N=0}

We consider the four-point function of scalar operators, $\f_Q$,
with dilation weight $Q$. From~\eq{4ptexp} this can be expanded in
the form

\be <\f_Q\f_Q\f_Q\f_Q>=g_{12}^Q g_{34}^Q \sum_{\cR} C_{\cR}
S_{\cR}(Z) \ee

where we have omitted the factor $\mbox{det}^p$ because this can
be absorbed into the expression for the Schur polynomial (since
$\mbox{det}^p(Z) S_{ab}(Z) = S_{a,b+p}(Z)$). Using~\eq{f2} we can
rewrite this in terms of  a function of two variables:

\be
 <\f_Q\f_Q\f_Q\f_Q>=g_{12}^Q g_{34}^Q
 G(x_1,x_2)
 \ee

 where

 \be
  G(x_1,x_2):= \sum_{a=0}^{\infty} \sum_b C_{ab} (x_1x_2)^b\  {x_1^{a+1}-x_2^{a+1} \over x_1-x_2}
 \ee

 and where we have written $C_{\cR_{ab}}=C_{ab}$. Note that the sum
 for $b$ will be over some countable set of real numbers,
 corresponding to the values of $b$ for which $C_{ab}$ is non-zero.

\subsection{Space-time independent four-point functions}\la{internal}

It is interesting to consider this formalism for the supergroup
$GL(0|n).$ This is equivalent to considering the four-point
function of spacetime independent $SU(2n)$ tensors. The operator
$A_Q$ is equivalent to a spacetime independent tensor carrying the
$SU(2n)$ representation with Dynkin labels $\left[0,\dots, 0,Q,0
\dots,0\right]$ where there are $n-1$ 0s on each side of the $Q$.

First consider $n=1$ in which case $Z=Y$. $A_Q$ is equivalent to a
space-time independent tensor carrying the $Q+1$ dimensional
representation of $SU(2)$. For example, if $Q=1$, then $A_Q(y)=A_1
+A_2 y$ is equivalent to the tensor $A_i$.

 We can again use~\eq{4ptexp}
to write the four-point function as

\be <QQQQ>=g_{12}^Q g_{34}^Q \sum_{p} C_{p}  \ Y^{-p}. \ee

Here we have used the fact that in this case
$S_{\cR_p}(Z)=\mbox{det}^p(Z)=\sd(Z)^{-p}$ where $\cR_p$ s given
by a Young tableau with one row of $p$ boxes. The latter identity
is simply the definition of superdeterminant for the group
$GL(0|n)$. We therefore omit $S_{\cR}(Z)$ but allow $p$ to take
positive and negative values.

Now $A_Q$ is analytic in the variables $Y_i$ and so the right hand
side must be also. Using the fact that $Y=y_{12}y^{-1}_{23}y_{34}
y^{-1}_{41}$, we find that the powers of $y_{ij}$ and
corresponding conditions are

\be \ba{rcl} y_{12}^Q y_{12}^{-p} & \quad \Rightarrow \quad & p \leq Q\\
    y_{13}^p  & \quad \Rightarrow \quad & p \geq 0 \ea \ee

and so there are $Q+1$ terms. This is precisely what one gets by
contracting $SU(2)$ indices.

We now consider $n=2$ corresponding to $SU(4)$ representations. A
field $A_Q$ is equivalent to an $SU(4)$ tensor with Dynkin labels
$[0,Q,0]$. If we consider the four-point function of four of these
we have

 \be <QQQQ>=(g_{12}g_{34})^Q \sum_{\cR} C_{\cR} \
 \ S_{\cR}(Z). \ee

 Here we omit the term $\sdet (Z)^{p}$ and instead consider $GL(2)$ Schur
 polynomials~\eq{schur}

 \be
 S_{cd}(Z)=(Y_1 Y_2)^{-d} \left({{Y_1}^{c+1}-{Y_2}^{c+1} \over Y_1-Y_2}
 \right)
 \ee

  with positive and negative values for $d$. Here we have let
  $Z \sim \diag(Y_1,Y_2)$. Now $Z=y_{21}y_{13}^{-1}y_{34}y_{42}^{-1}$ and $Y_1 Y_2=\sdet
Z^{-1}=(y_{12}^2 y_{34}^2)(y_{13}^2y_{24}^2)^{-1}$ so that

\be (Y_1Y_2)^{-d}=(y_{12}^2 y_{34}^2)^{-d}(y_{13}^2 y_{24}^2)^d
\ee

On the other hand, the second factor in the sum is always a
positive power of $Z$. So the only poles here will come from
inverting $y_{13}$ and $y_{42}$. Thus this factor will contain a
factor of $(y_{13}^2y_{24}^2)^{-c}$.

 Analyticity in the variables $y_{12}$ or $y_{34}$ therefore implies
that $Q-d\geq 0$ while analyticity in $y_{13}$ or $y_{24}$ implies
that $d-c\geq 0$. So we conclude that $0\leq c\leq d\leq Q$. This
gives a total of $\half (Q+1)(Q+2)$
 terms (eg 6 terms if $Q=2$ corresponding to the four point function of
four
 tensors in the 20' representation of $SU(4)$). This is precisely the
number
 of different irreducible representations
 one obtains in $[0,Q,0] \xz
 [0,Q,0]$ which is what one would expect. The formalism therefore
provides a nice way to study the internal group which has some
relevance to the leading term of the four-point functions of CPOs
as we shall see later. However, the real power of the
 formalism occurs when one combines the spacetime and internal parts.

\subsection{Four-point function in $N=2$}

\subsection{$<2222>$}

We now consider the four-point function of four CPOs of charge two
in $N=2$ SCFT, reproducing results in~\cite{Dolan:2001tt} (note,
however, that we write down
 the
complete four-point function for four charge 2 multiplets, whereas
previously results have been only for the first term in theta.) We
expand this in the form~\eq{4ptexp}:

\be <2222>=g_{12}^2 g_{34}^2 \sum_{p,\cR} C_{p,\cR}  \ \sdet
(Z)^{p} \ S_{\cR}(Z) \ee

where $g_{12}={\hat y_{12}\over x_{12}^2}$ with $\hat
y_{12}=y_{12}-\p_{12}x^{-1}_{12} \l_{12}$. We can restrict the
$\cR$s to be given by Young tableaux with only two rows since a
tableau with more than two rows can be replaced by one with two
rows if we include an appropriate power of $\sdet (Z)$.)

We now consider the constraints imposed by analyticity. In the
$(12)$ channel we have a factor $(y_{12})^{-p}$ coming from the
superdeterminant and a factor $(y_{12})^2$ from the propagator, so
that analyticity in $y_{12}$ implies that $p\leq 2$. The same
result is obtained from  the $(34)$ channel .  The $S_{\cR}$
factor will only contribute possible singularities in the $(13)$
and $(24)$ channels. From the symmetry properties of the
representation $\cR$ it is clear that we can have no more than $r$
factors of $y_{13}^{-1}$ (or $y_{24}^{-1}$), where $r$ is the
number of rows in the tableau. In the $(13)$ channel we have a
factor of $y_{13}^p$ from the superdeterminant, so analyticity in
this channel implies that $p\geq r$. The same result is obtained
for the channel $(24)$.  Since $r=0,1,2$ is the number of rows of
a representation we see that $p=2$ for long representations
$\cR_{ab}$, $p=2$ or $1$ for the representations $\cR_{a}$ and
$p=0,1,2$ for the trivial representation.

We can therefore write

\be <2222>=g_{13}^2 g_{24}^2 \left( \sum_{a=0}^{\infty} \sum_{b>1}
C_{ab}
  \ S_{ab} (Z) + \sum_{a=-1}^{\infty} C_{a} S_a(Z) + {Y \over
  X_1X_2}\sum_{a=-2}^{\infty} C'_{a} S_a(Z) \right) \la{4pt} \ee

where we have absorbed the factor $(\sdet Z)^{2}$ into the
prefactor and where we have defined $C_{ab}:=C_{2,\cR_{ab}},\
C_a:= C_{2,\cR_{a}},\  C'_a:=C_{1,\cR_{a}}$ and
$C'_{-2}=C_{0,\cR_{0}}$. We have also used the fact that if we put
$a=-2$ into the formula~\eq{S_a} for $S_a$ we find $S_{-2}={Y
\over X_1X_2}$. Note that by the sum over $b>1$ we mean that the sum
is over some countable set of real numbers greater than one. In the
free theory, the sum will be over integers as illustrated in
section~\ref{free}, but in the interacting theory the sum will in
general be over non-integers corresponding to the presence of anomalous
dimensions.

Using
 \eqs{Ff}{Gg} we can write this expression in terms of two functions of one
variable $f(X_1),f'(X_1)$ and a function of two variables
$G(X_1,X_2)$

\be <2222>=g_{13}^2 g_{24}^2 \left( \cG[G] + \cF[f] + {Y \over
X_1X_2} \cF[f'] \right) \la{expr} \ee

where

\be f(X_1)= \sum_{a=-1}^{\infty} C_a X_1^{a+1} \quad f'(X_1)=
\sum_{a=-2}^{\infty}
 C'_a
X_1^{a+1} \quad G(X_1,X_2)=\sum_{a=0}^{\infty} \sum_{b>1}C_{ab}
 (X_1X_2)^{b-2}\ {X_1^{a+1}-X_2^{a+1} \over X_1-X_2}.
\ee

We therefore arrive at a compact formula for the fully
supersymmetric $N=2$ four-point function of four charge 2
operators in terms of two functions of one variable $f,f'$ and a
function of two variables $G$.

There are two functions of one variable since there are two
sequences of short representations in the four-point function
(corresponding to $p=1$ and $p=2$) and there is one function of
two variables since there is only one sequence of long
representations. This is a general feature, for each sequence of
short representations one gets a function of one variable, and for
each sequence of long representations one gets a function of two
variables.

The two-variable contribution to the correlator can be written,
using \eq{Gg} in the form

\be <2222>\sim (g_{12}g_{34})^2 S_{02}\  G(X_1,X_2) \ee

where we recall that $S_{02}=X_1X_2-Y(X_1+X_2) +Y^2$.

One can, if one prefers, absorb one of the single variable
functions into the function of two variables, $G$ by
using~\eq{red} and by allowing terms with $b=1$ in $G$.
Equation~\eq{red} implies that

\be {Y \over X_1X_2}\cF[h(X_1)]-\cF\left[{{h(X_1)\over
X_1}}\right] +\cG\left[{
 {h(X_1)-h(X_2)\over (X_1X_2)(X_1-X_2)}}\right]=0\la{fnl}
\ee

for any function of one variable $h$, and since the functionals
$\cF,\cG$ are linear, the expression for the four point
function~\eq{expr} is invariant under

\be f'\rightarrow f' - h \qquad f \rightarrow f + {h\over X_1}
\qquad G \rightarrow G - {h(X_1)-h(X_2)\over X_1X_2(X_1-X_2)} \ee

So in particular, with the choice $h=f'$ we can absorb  $f'$
entirely into $G(X_1,X_2)$ and write the entire four-point
function in terms of a single function of one variable and a
function of two variables at the expense of allowing lower powers
of $X_1,X_2$ in $G$.

From~\eq{eq:18} we can see that the coefficients $C_a,C'_a$ are
related to the three-point functions of two $T$s and a protected
operator. In~\cite{Heslop:2001gp} we showed that all such
three-point functions receive no corrections to their free-field
values. Therefore one expects the functions $f,f'$ to take their
free-field values.

\subsection{$<QQQQ>$}

It is straightforward to extend this to the case of CPOs of
arbitrary charge $Q$.

\be <QQQQ>=(g_{12}g_{34})^Q \sum_{p,\cR} C_{p,\cR} \ \sdet
(Z)^{-p} \ S_{\cR}(Z).\la{Q} \ee

Analyticity implies that $r\leq p\leq Q$, where $r$ is the number
of rows in the Young tableau which can be at most two.

We have

\be \ba{rl} <QQQQ>=(g_{13} g_{24})^Q &\left( \sum_{p=2}^Q
\sum_{a=0}^{\infty}
  \sum_{b>1}\left( {Y \over X_1X_2} \right)^{Q-p}  C_{pab} S_{ab}(Z)
\right. \\
& +\left. \sum_{p=1}^Q \sum_{a=-1}^{\infty}\left( {Y \over X_1X_2}
\right)^{Q-p}
 C_{pa}
S_{a}(Z) + C \left( {Y \over X_1X_2} \right)^Q \right).\ea \ee

It is convenient to rewrite the last term as

\be C \left( {Y \over X_1X_2} \right)^Q=\left( {Y \over X_1X_2}
\right)^{Q-1} C_{1,
 -2}
S_{-2}(Z). \ee

This four-point function can then be rewritten in terms of $Q-1$
functions of two variables, $G_p(X_1,X_2)$ and $Q$ functions of of
one variable, $f_p(X_1)$, as

\be <QQQQ>=(g_{13} g_{24})^Q  \left( \sum_{p=2}^Q \left( {Y \over
X_1X_2}
  \right)^{Q-p} \cG[G_p] + \sum_{p=1}^Q \left( {Y \over X_1X_2}
\right)^{Q-p}
  \cF[f_p] \right)
\ee

where

\bea
f_p &=& \sum_{a=-1}^{\infty}C_{pa}X_1^{a+1} \qquad p=2\cdots Q\\
f_1&=&\sum_{a=-2}^{\infty}C_{1a}X_1^{a+1} \\
G_p(X_1,X_2)&=&\sum_{a=0}^{\infty} \sum_{b>1}C_{pab}
 (X_1X_2)^{b-2}\ {X_1^{a+1}-X_2^{a+1}\over X_1-X_2}.
\eea

Using \eq{Gg} we can rewrite the two-variable contribution in the
form

\be <QQQQ>_2= (g_{13}g_{24})^Q S_{02} \sum_{p=2}^Q G_p(X_1,X_2)
Y^{Q-p} \ee

where we have absorbed some factors of $X_1$ and $X_2$ into the
$Q-1$ $G_p$s.

As in the $<2222>$ case we could use~\eq{red} to remove some of
the one-variable functions, $f_p$ by subsuming them into the two
variable functions. By repeated application of~\eq{fnl} one
obtains the relation

\be \left({Y \over X_1X_2}\right)^p \cF[h(X_1)]=\cF\left[
{h(X_1)\over
 X_1^p}\right] -\cG\left[{ {h'(X_1)-h'(X_2)\over (X_1X_2)(X_1-X_2)}}\right]
\ee

where

\be h'(X_1):= (1 + {1\over X_1} + \cdots + {1\over
X_1^{p-1}})h(X_1) \ee

which  can be used to absorb the functions $f_p,\ p=1\dots Q-1$
into $f_Q$ and
 $G_p$. This will leave us with only one function of one variable, but
at the
 expense of allowing powers of $X_1^{-p}$ in the remaining functions
$f_Q$ and
 $G_p$.

Since the functions of one variable will take their free theory
values (that is, they are non-renormalised) it does not seem to be
particularly advantageous to absorb the single variable functions
in this way.

\subsection{Four-point functions in $N=4$}

\subsubsection{Charge 2}

We can  make a similar construction in $N=4$ for the four-point
function of four CPOs. We begin with the simplest case of charge
$Q$. For this case the operator $A_2$ is the supercurrent
supermultiplet $T$. We first expand the four point function as

\be <T T T T>=(g_{12} g_{34})^2 \sum_{p,\cR} C_{p,\cR}  \
\sd(Z)^{p} \ S_\cR(Z) \ee

and then consider analyticity in the various internal coordinates.
Since we are now in $N=4$ the representations $\cR$ that occur in
the above formula are representations of $GL(2|2)$ . The most
general such representation (up to multiplication by a
superdeterminant) which can occur can be specified by a Young
tableau of the form

\be
\begin{picture}(100,60)(0,-40) \put(0,0){$\overbrace{\square
\! \square  \cdots \square \! \square}^b \! \!
 \overbrace{\square \! \square \cdots  \square \! \square}^a $}
\put(0,-7){$ \square \! \square  \cdots \square \! \square  $}
\put(-14.7,-30.4){$\left\{ \ba{c} \square \\ \vdots \\ \square \ea
\right. $} \put(-20,-30){$\scriptstyle c$}
\end{picture}
\ee

As in the previous examples we need only look at channels $(12)$
and $(13)$ in order to obtain the restrictions due to analyticity.
The superdeterminant contributes a factor $(y_{12}^2)^{-p}
(y_{13}^2)^{p}$ so that analyticity in the $(12)$ channel implies
that $p\leq 2$.  The maximum number of powers of $y_{13}^{-1}$
that can arise from the representation $\cR$ described by the
above tableau will occur when all the boxes in the first two
columns are filled with $y_{13}^{-1}$s.  There   are $r-c$ boxes
with two $y_{13}^{-1}$s giving a factor $(y^2_{13})^{c-r}$ and $c$
boxes with $y_{13}^{-1}$s in a symmetrised combination. This gives
a factor of $(y_{13}^{-1})^c$ and this in turn gives rise to a
factor of $(y^2_{13})^{-c}$. Analyticity in the $(13)$ channel
therefore yields the constraint $p\geq r$. Here $r$ is the total
number of rows in the tableau, so $p=2$ for representations with
two rows, $p=1$ or 2 for representations $\cR_{a,0}$
 with only 1 row, and
$p=0,1,2$ for the trivial
 representation $\cR_{0}$.
 So the four-point function has the form

\be \renewcommand{\arraystretch}{1.5} \ba{rl} <T T T T>=g_{13}^2
g_{24}^2 &\left(\sum_{a=0}^{\infty} \sum_{b>1} C_{2ab0} \
S_{ab0}(Z) + \sum_{a=-1}^{\infty}C_{2a1} S_{a1}(Z) +
\sum_{a=-2}^{\infty} C_{2a0} S_{a0}(Z)\right. \\
&+ \left. \left({Y_1 Y_2 \over X_1X_2}\right)\sum_{a=-2}^{\infty}
C_{1a0} S_{a0}(Z) + C \right). \ea \ee

In this expression we have used the fact that $S_{-2,0}=1-Y_1 Y_2 /
X_1X_2=S_0-(Y_1 Y_2 / X_1X_2)S_0$ to reexpress some terms. As in
previous cases, the sum over $b$ is a sum over some countable set of
real numbers.

This can then be written as

\be \renewcommand{\arraystretch}{1.5} \ba{rl} <TTTT>=g_{13}^2
g_{24}^2 &\left( \cG_{ab0}[G_0] + \cF_1[f] + \cF_0[g]+
  \left({Y_1 Y_2 \over X_1X_2}\right) \cF_0[h] + C  \right)
\ea \ee

with

\bea G_0(X_1,X_2)&=&\sum_{a=0}^{\infty} \sum_{b>1} C_{2ab0}
(X_1X_2)^{b-2}\ {X_1^{a+1}-X_2^{a+1} \over X_1-X_2}\\
f(X_1)&=& \sum_{a=-1}^{\infty}C_{2a1} X_1^a\\
g(X_1)&=& \sum_{a=-2}^{\infty}C_{2a0} X_1^a\\
h(X_1)&=& \sum_{a=-2}^{\infty}C_{1a0} X_1^a. \eea

We have rewritten the four point function in terms of one function
of two variables, $G(X_1,X_2)$, three functions of one variable,
$f(X_1), g(X_1),
 h(X_1)$ and a
constant, $C$. As in the $N=2$ case one can eliminate a function
of one variable
 using~\eq{red2} by
including terms of order $1/(X_1X_2)$ in $G(X_1,X_2)$.

\subsection{$<QQQQ>$}

The above formula can easily be generalised to the four-point
function of charge $Q$ CPOs in analytic superspace

\be <QQQQ>=(g_{12}g_{34})^Q \sum_{p,\cR} C_{p,\cR}  \ \sd(Z)^{p} \
\str\cR(Z). \ee

The analyticity conditions become $p\leq Q$ (from the $(12)$
channel) and $p\geq r$ (from the $(13)$ channel), so that the
maximum number of rows that is allowed is now $Q$.

 We therefore obtain

\be \ba{rcll}
 <QQQQ>&=&g_{13}^Q g_{24}^Q &\left( \sum_{p=2}^{Q} \sum_{c=0}^{p-2}
   \sum_{a=0}^{\infty} \sum_{b>1} \left( {Y_1 Y_2 \over X_1 X_2}
\right)^{Q-p}
   C_{pabc} S_{abc}\right. \\
&&&\left. + \sum_{p=1}^{Q}
 \sum_{c=0}^{p-1} \sum_{a=-2}^{\infty} \sum_{b>1}\left( {Y_1 Y_2 \over X_1
     X_2} \right)^{Q-p}
C_{pac}S_{ac}
 + C \right)\\
\\
&=& g_{13}^Q g_{24}^Q &\left( \sum_{p=2}^{Q} \sum_{c=0}^{p-2}
\left( {Y_1 Y_2 \over X_1 X_2} \right)^{Q-p} \cG_c[G_{pc}]
 + \sum_{p=1}^{Q}
 \sum_{c=0}^{p-1} \left( {Y_1 Y_2 \over X_1 X_2} \right)^{Q-p}
\cF_c[f_{pc}] +  C \right)\\
\ea \ee

We see that  the four-point function can be expressed in terms of
$(1/2)Q(Q-1)$ functions of two variables, $G_{pc}$, $(1/2)Q(Q+1)$
functions of one variable, $f_{pc}$ and a constant, $C$.

We can once again use~\eq{red2} to absorb $(1/2)Q(Q-1)$ of the
one-variable functions into the two variable functions to leave us
with $Q$ functions of one variable, $(1/2)Q(Q-1)$ functions of two
variables and a constant.

We can rewrite the two-variable contribution in a slightly more
explicit form by using \eq{eq:52}.  We find

\be <QQQQ>_2= (g_{13}g_{24})^Q S_{020} \sum_{p=2}^Q
\sum_{c=0}^{p-2} G_{pc} \left({Y_1^{c+1}-Y_2^{c+1}\over Y_1-Y_2
}\right) (Y_1Y_2)^{Q-p} \ee

where $S_{020}$ is a universal factor which is given in
equation~\eq{S020} and where we have again absorbed some explicit
factors of $X_1$and $X_2$ into the functions $G_{pc}$. Comparing
this result with equation \eq{mainresult} in section 2, we see
that

\be F^Q=\sum_{p=2}^Q \sum_{c=0}^{p-2} G_{pc}
\left({Y_1^{c+1}-Y_2^{c+1}\over Y_1-Y_2 }\right) (Y_1Y_2)^{Q-p}
\ee

\section{Crossing symmetry}

In this section we shall discuss the restrictions that crossing
symmetry imposes on the four-point functions. We begin with the
$N=2$ case.

\subsection{$N=2$}

We recall that the two-variable contribution to the four-point
function is

\be <QQQQ>_2= (g_{13}g_{24})^Q S_{02} \sum_{p=2}^Q G_p(X_1,X_2)
Y^{Q-p} \ee

If we interchange points $1$ and $3$ we find that the prefactor
changes only by a sign  while $Z\mapsto (1-Z)$. If we interchange
points 1 and 4 $Z\mapsto Z^{-1}$ while the prefactor is multiplied
by $Y^{Q-2} (X_1 X_2)^{-Q-2}$. Now the number of independent
function is determined by the behaviour of the sum as a function
of $Y$. To determine this we can ignore the dependence of the $G$s
on $X_1,X_2$. The problem is therefore to find the number of
independent polynomials in $Y$ of degree $n:=Q-2$, $f(Y)$, such
that

\be
 f(Y)=f(1-Y)\qquad{\rm and}\qquad f(Y)=Y^n f(Y^{-1})
\ee

Starting from $Y$ we generate a sequence of six polynomials $Y,
(1-Y), Y^{n-1}, (1-Y)^{n-1}, Y(1-Y)^{n-1},(1-Y)Y^{n-1}$ provided
that $n\geq 5$. For $n\leq 5$ it is easy to see that this
procedure generates a basis of polynomials of degree $n$, so that
there is just one invariant in all of these cases. For $n>5$ we
need to consider  a second sequence which can be generated from
$Y^2$, and so forth. We therefore conclude that the number of
independent functions of two-variables which is needed in the
fully crossing symmetric four-point function of charge $Q$ CPOs is
$[1/6\, (Q-1)]$ where the square brackets denote the smallest
integer which is greater than or equal to the number inside.

\subsection{$N=4$}

For the $N=4$ case we start from the expression

\be <QQQQ>_2= (g_{13}g_{24})^Q S_{020} \sum_{p=2}^Q
\sum_{c=0}^{p-2} G_{pc} \left({Y_1^{c+1}-Y_2^{c+1}\over Y_1-Y_2
}\right) (Y_1Y_2)^{Q-p} \la{eq:102}\ee

Under the interchange of points 1 and 3 the prefactor here is
invariant while $Z\mapsto (1-Z)$, while under the interchange of
points 1 and 4 the prefactor is multiplied by
$(Y_1Y_2)^{Q-2}(X_1X_2)^{-(Q+2)}$.  Now the sum in~\eq{eq:102} can
be rewritten as a polynomial in the variables $Y_1+Y_2$ and
$Y_1Y_2$ of degree $n=(Q-2)$, so we again need to decide how many
such polynomials there are which are invariant under the
transformations generated by the crossing symmetries. To do this
it is convenient to consider instead polynomials of degree $n$ in
the variables $y:Y_1Y_2$ and $z:=(1-Y_1)(1-Y_2)$. If we denote
such a polynomial by $f(y,z)$ then the invariance conditions are

\be
 f(y,z)=f(z,y)\qquad{\rm and}\qquad f(y,z)=y^n f(y^{-1},y^{-1}z).
\ee

The basis elements $w^{rs}:=y^r z^s$ behave as

\bea
w^{rs} &\mapsto & w^{sr}\\
w^{rs} &\mapsto & w^{n-(r+s),s} \eea

under the crossing transformations. These basis elements can be
related to each other in various chains under these
transformations. For example,

\be 1\mapsto y^n \mapsto z^n \ee

and

\be y\mapsto y^{n-1}\mapsto z^{n-1}\mapsto yz^{n-1} \mapsto
y^{n-1}z \mapsto z \ee

In fact,  the chains of monomials generated in this fashion either
have length 3 or length 6 apart from some exceptional chains of
length 1 which occur when $r=s=n/3$.

Using this information one can show that the number, $N_Q$,  of
invariant polynomials of this type is given by the following
formula:

\be \ba{ll}
Q-2=0\ {\rm mod}\ 6: &  N_Q=\frac{(Q-2)(Q+4)}{12 }+ 1 \\
&\\
Q-2=1\ {\rm mod}\ 6:  &  N_Q=\frac{(Q+3)(Q-1)}{12}\\
&\\
Q-2=2\ {\rm mod}\ 6: &  N_Q=\frac{(Q+2)Q}{12}\\
&\\
Q-2=3\ {\rm mod}\ 6: &  N_Q=\frac{(Q+1)2}{12}\\
&\\
Q-2=4\ {\rm mod}\ 6: &  N_Q=\frac{Q(Q+2)}{12}\\
&\\
Q-2=5\ {\rm mod}\ 6: &  N_Q=\frac{(Q-1)(Q+3)}{12} \ea \ee

\section{Free four-point functions}\la{free}

As we have stated above, it is expected that the single-variable
functions that occur in the four-point functions will take the
same forms as they do in the free theory. In this section we
express these free functions in terms of the variables we have
been using for the case of charge 2 operators.

The free four-point function of four charge two operators, in both
$N=2$ and $N=4$, has the form

\be <2222>=A(g_{12}^2 g_{34}^2 + g_{13}^2 g_{24}^2 + g_{14}^2
g_{23}^2)
      +B(g_{12}g_{34}g_{13}g_{24}+ g_{12}g_{34}g_{14}g_{23}+ g_{13}g_{24}g_{14}g_{23} )
\ee

which can be rewritten as

\be \ba{rl}
<2222>= g_{13}^2 g_{24}^2& \left( A(1+\sdet (Z)^{-2} + \sd(1-Z)^{-2})\right.\\
     & +B\left.(\sdet (Z)^{-1} + \sdet (1-Z)^{-1} + \sdet (Z)^{-1}\sdet (1-Z)^{-1})\right).
\la{free4pt} \ea \ee

One must now distinguish between $N=2$ and $N=4$. For $N=2$ one
can show that

\bea \sdet (1-Z)^{-2}&=&\sum_{b=1}^{\infty} \sum_{a=0}^{\infty}
(a+1)S_{a,b+1} +
             \sum_{a=0}^{\infty} (a+1)S_{a-1}\\
\sdet (1-Z)^{-1}&=&\sum_{a=0}^{\infty} S_{a-1} \eea

and by inserting these expressions into~\eq{free4pt} one can
rewrite this in the form~\eq{4pt} with

\bea
G(X_1,X_2)&=&{A \over (1-X_1)^2(1-X_2)^2}\\
f(X)&=&A+A{1\over(1-X)^{2}}+ B {1 \over (1-X)}\\
f'(X)&=& {A \over X} + B\left(1+{1 \over 1-X}\right). \eea

Then using the invariance following from \eq{fnl} we can absorb
$f'$ into $f$ and $G$ to give

\bea {G(X_1,X_2)}&=&A\left({X_1-X_2 \over (1-X_1)^2(1-X_2)^2}
+ {1 \over X_1^2 X_2^2}\right) - {B \over X_1X_2(1-X_1)(1-X_2)}\\
f(X)&=&A\left( {1 \over (1-X)^2}+ {1 \over X^2} + 1 \right) + B
\left({1 \over (1-X)}+ {1 \over X} +
{1 \over X(1-X)} \right)\\
f'(X)&=&0. \eea

In this form the crossing symmetry is more apparent.

In $N=4$ we have

\bea \sdet(1-Z)^{-2}&=&\sum_{b=2}^{\infty} \sum_{a=0}^{\infty}
(a+1)S_{ab0} +
              \sum_{a=0}^{\infty} \left((a+1)S_{a-1,1} + (a+1)S_{a-2,0} \right)\\
\sdet (1-Z)^{-1}&=&\sum_{a=0}^{\infty} S_{a-2,0} \eea

and so \eq{free4pt} can be rewritten in terms of the functions
$G_0,f,g,h$ and the constant $C$ as

\be <TTTT>=g_{13}^2 g_{24}^2 \left( \cG_{ab0}[G_0] + \cF_1[f] +
\cF_0[g]+
  \left({Y_1 Y_2 \over X_1X_2}\right) \cF_0[h] + C \right)
\ee

where

\bea
{G_0}&=&{A \over (1-X_1)^2(1-X_2)^2}\\
f(X)&=&{A \over X(1-X)^2}\\
g(X)&=&{-(2A+2B)\over X^2} + {A\over X^2(1-X)^2}+ {B \over X(1-X)}\\
h(X)&=& {-A \over X^2} + {B \over X(1-X)}\\
C&=& 3A +3B. \eea

In a similar fashion to the $N=2$ case we can absorb the function
$h$ to rewrite this in the form

\bea {G_0}&=&A \left( { 1\over (1-X_1)^2(1-X_2)^2} +
{1 \over (X_1 X_2)^2} \right) + {B\over (1-X_1)(1-X_2)X_1X_2}\\
f(X)&=&{1 \over X}\left(A \left({1\over (1-X)^2}+ {1\over X^2}\right) - {B \over X(1-X)}\right)\\
g(X)&=&{-(2A+2B)\over X^2} + {A\over X^2(1-X)^2}+ {B \over X(1-X)}\\
h(X)&=& 0\\
C&=& 3A +3B. \eea

\section{Conclusions}

We have considered in this paper the analysis of four-point
superconformal invariants by means of the Schur polynomials. The
constraints imposed on the four-point functions by superconformal
symmetry come from the Ward identities together with analyticity
in the internal coordinates. These conditions can be solved
straightforwardly in terms of a number of functions of two
variables and a number of single-variable functions, the
particular numbers depending on the charge of the operators. The
single-variable functions are due to the contributions of
protected operators in the OPE and are therefore expected to take
their free theory forms as they arise from three-point functions
of two CPOs and the protected operator in question.  This is
partial non-renormalisation.The fact that these three-point
functions are themselves non-renormalised can be proved by means
of the reduction formula which relates an $n$-point to an
$(n+1)$-point function with an integrated insertion of the
stress-energy tensor. It therefore appears that the structure of
such four-point functions in $N=4$ SYM is a corollary of the
existence of protected operators and the non-renormalisation
theorem for three-point functions of them.

We have concentrated in this paper on four-point functions of
identical operators but it should be possible to generalise the
theory to four-point functions of arbitrary CPOs. It would be
interesting to extend the formalism further to include other
series C operators, for example, 1/4 BPS superfields. Although
this would be technically more complicated, one might expect to
find similar features because the short operators in the OPE would
again give non-renormalised contributions.

Although we have focused on the $N=4$ theory, the formalism is
applicable to $N=2$ and even $N=0$ as we have seen. We expect that
it can be applied to other cases such as the $(2,0) D=6$
superconformal theory \cite{hst}, which also has a  harmonic
superspace formulation \cite{Howe:1998jw}. Indeed, the four-point function
of supercurrents has been investigated and compared with the relevant
supergravity in \cite{Arutyunov:2002ff}. The results are very similar to the
four-dimensional case although they were obtained without the
use of the reduction formula.

\section*{Acknowledgements}

This research was supported in part by  PPARC SPG grant 613,
PPARC Rolling Grant 451 and EU grant HPRN 2000-00122.
P. Heslop acknowledges financial support from the RTN European
Program HPRN-CT-2000-00148 ``Physics Across the Present Energy
Frontier: Probing the Origin of Mass.''

We are grateful to Emery Sokatchev for many interesting
discussions. The results for the number of independent functions
of two variables and the partial non-renormalisation of $N=4$
four-point functions presented here have also been obtained (for
the leading component fields) in the CPO multiplets by the use of
the reduction formula in superspace and also in a supergravity
calculation, the two calculations being in agreement with each other.
These results will appear in a forthcoming paper
\cite{ae}.

\providecommand{\href}[2]{#2}\begingroup\raggedright
\endgroup

\end{document}